\documentstyle[12pt,a4]{article}
\begin{document}\bigskip\bigskip
\begin{center}
{\bf \Large{SPINNING PARTICLES IN TAUB-NUT SPACE}}
\end{center}
\vskip 1.0truecm
\centerline{{\bf\large
{Diana Vaman\footnote{E-mail 
address:~~~~dvaman@insti.physics.sunysb.edu}~~    
{\it and}~~ {\bf\large Mihai Visinescu\footnote
{E-mail address:~~~ mvisin@theor1.ifa.ro}}}}}
\vskip5mm
\centerline{Department of Theoretical Physics}
\centerline{Institute of Atomic Physics, P.O.Box MG-6, Magurele,}
\centerline{Bucharest, Romania}                                      
\vskip 2cm

\bigskip \nopagebreak \begin{abstract}
\noindent
The geodesic motion of pseudo-classical spinning particles in
Euclidean Taub-NUT space is analysed. The constants of motion
are expressed in terms of Killing-Yano tensors. Some previous
results from the literature are corrected.
\end{abstract}

\vskip 1cm
PACS number(s): ~~~~04.20.Jb,~~02.40.-K
\vskip 1cm

The configuration space of spinning particles (spinning space)
is an supersymmetric extension of an ordinary Riemannian
manifold, parametrized by local coordinates $\{x^\mu \}$, to a
graded manifold parametrized by local coordinates $\{ x^\mu ,
\psi^\mu \}$, with the first set of variables being Grassmann
even (commuting) and the second set, Grassmann odd
(anticommuting). The equation of motion of a spinning particle
on a geodesic is derived from the action:
\begin{equation}
S=\int_{a}^{b}d\tau \left(\,{1\over 2}\,g_{\mu \nu}(x)\,\dot{x}^\mu 
\,\dot{x}^\nu\, +\, {i\over 2}\, g_{\mu \nu}(x)\,\psi^\mu \,
{D\psi^\nu\over D\tau} \right). 
\end{equation}

The corresponding world-line Hamiltonian is given by:
\begin{equation}
H=\frac1 2 g^{\mu\nu}\Pi_\mu \Pi_\nu
\end{equation}
where $\Pi_\mu = g_{\mu\nu}\dot{x}^\nu$ is the covariant 
momentum. 

For any constant of motion ${\cal J}(x,\Pi,\psi)$, the bracket
with $H$ vanishes $\{ H,{\cal J} \} = 0$.

If we expand ${\cal J}(x,\Pi,\psi)$ in a power series in the
covariant momentum 
\begin{equation}
{\cal J}=\sum_{n=0}^{\infty}\frac{1}{n!}{\cal J}^{(n)\mu_1
\dots\mu_n}(x,\psi) \Pi_{\mu_1}\dots\Pi_{\mu_n}
\end{equation}
then the bracket $\{ H , {\cal J}\}$ vanishes for
arbitrary $\Pi_\mu$ if and only if the components of ${\cal J}$ 
satisfy the generalized Killing equations $[1,2]$:
\begin{eqnarray}
\nonumber
{\cal J}^{(n)}_{(\mu_1\dots\mu_n;\mu_{n+1})} + \frac{\partial
{\cal J}^{(n)}_{(\mu_1 \dots\mu_n}}{\partial \psi^\sigma}
\Gamma^\sigma_{\mu_{n+1})\lambda} \psi^\lambda \\
=\frac{i}{2}\psi^\rho \psi^\sigma R_{\rho\sigma\nu(\mu_{n+1}}
{{\cal J}^{(n+1)\nu}}_{\mu_1 \dots \mu_n)}.
\end{eqnarray}

In general, the symmetries of a spinning-particle model can be 
divided into two classes [1].  First there are four independent 
{\it generic} symmetries  which exist for any spinning particle 
model (1) and {\it non-generic} ones, which depend on the specific
background space considered. 

To the first class belong: proper-time translations generated 
by the hamiltonian $H$; supersymmetry generated by the supercharge
\begin{equation}
Q_0=\Pi_\mu\,\psi^\mu,
\end{equation}
and furthermore chiral and dual supersymmetry, generated
respectively by the chiral charge
\begin{equation}
\Gamma_* = \frac{i^{[\frac{d}{2}]}}{d!}\sqrt{g}\epsilon_{\mu_1 
\dots \mu_d} \psi^{\mu_1} \dots \psi^{\mu_d},
\end{equation}
and dual supercharge
\begin{equation}
Q^* =i\{ \Gamma_* , Q_0 \}=
\frac{i^{[\frac{d}{2}]}}{(d-1)!}\sqrt{g}\epsilon_{\mu_1 \dots 
\mu_d} \Pi^{\mu_1}\psi^{\mu_2} \dots \psi^{\mu_d}
\end{equation}
where $d$ is the dimension of space-time.

In a recent paper [3] we have pointed out the significant role
of the Killing-Yano tensors in generating solutions of the
generalized Killing equations (4).

The aim of this paper is to apply the general results to the
case of the four dimensional Euclidean Taub-NUT manifold.
The motivation of this paper is twofold. First of all, in the
Taub-NUT geometry there are known to exist four Killing-Yano
tensors [4]. From this point of view, the spinning Taub-NUT
space is an exceedingly interesting space to exemplify the
effective construction of all conserved quantities in terms of
geometric ones, namely Killing-Yano tensors. On the other hand,
the Taub-NUT geometry is involved in many modern studies in
physics. For example the Kaluza-Klein 
monopole of Gross and Perry [5] and of Sorkin [6] was obtained by 
embedding the Taub-NUT gravitational instanton into five-dimensional 
Kaluza-Klein theory. Remarkably, the same object has re-emerged in 
the study of monopole scattering. In the long distance limit, 
neglecting radiation, the relative motion of the BPS monopoles is 
described by the geodesics of this space [7,8]. The dynamics of 
well-separated monopoles is completely soluble and has a Kepler 
type symmetry [4, 9-11].

Therefore, having in mind the importance of the geodesic motion
in Taub-NUT space, we extend the study to the spinning Taub-NUT
space, correcting some inaccurate relations present in the
literature. 

In a special choice of coordinates the Euclidean Taub-NUT metric
takes the form 
\begin{equation}
ds^2 = V(r)\left(dr^2+r^2 d\theta^2+r^2\sin^2\theta
 d\varphi^2\right) + 16m^2 V^{-1}(r)(d\chi+\cos\theta
 d\varphi)^2
\end{equation}
with $V(r)=1 + \frac{4m}{r}$.

There are four Killing vectors
\begin{equation}
D_A=R_{A}^\mu\,\partial_\mu,~~~~A=0,\cdots ,3
\end{equation}
corresponding to the invariance of the metric (8) under spatial 
rotations $(A=1,2,3)$ and $\chi$ translations $(A=0)$ [12].

In the purely bosonic case these invariances would correspond to 
conservation of angular momentum and ``relative electric charge'':
\begin{equation}
\vec{j}=\vec{r}\times\vec{p}\,+\,q\,{\vec{r}\over r} 
\end{equation}
\begin{equation}
q=16m^2\,V(r)\,(\dot\chi+\cos\theta\,\dot\varphi)
\end{equation}
where $\vec{p}={1\over V(r)}\dot{\vec{r}}$ is the 
``mechanical momentum'' which is only part of the momentum
canonically conjugate to $\vec{r}$.  

In the Taub-NUT case there is a conserved vector, analogous to
the Runge-Lenz vector of the Kepler type problem:
\begin{equation}
\vec K= \frac1 2 {\vec K}_{\mu\nu} \Pi^\mu \Pi^\nu= \vec
p \times \vec j + \left( \frac{q^2}{4m}-4mE\right)\frac{\vec r}{r}
\end{equation}
where the conserved energy, from Eq.(2) is
\begin{equation}
E=\frac1 2 g^{\mu\nu} \Pi_\mu \Pi_\nu=\frac1 2 V^{-1}(r)\left[
{\dot{\vec r}}^{~2} + \left(\frac{q}{4m}\right)^2 \right].
\end{equation}

A tensor  $f_{\mu_1 \dots\mu_r}$ 
is called a Killing-Yano tensor of valence $r$ if it 
is totally antisymmetric and satisfies the equation 
\begin{equation}
f_{\mu_1 \dots\mu_{r-1}(\mu_{r};\lambda)} = 0.
\end{equation}

Its existence is equivalent to the existence of a 
supersymmetry for the spinning space, with a supercharge 
\begin{equation}
Q_f = f_{\mu_1 \dots\mu_r}\Pi^{\mu_1}\psi^{\mu_2}\dots \psi^{\mu_r} 
+ \frac{i}{r+1}(-1)^{r+1}f_{[\mu_1 \dots
\mu_r;\mu_{r+1}]}\psi^{\mu_1}\dots \psi^{\mu_{r+1}}
\end{equation}
which anticommutes with $Q_0$ [2,3,12,13].

The Killing-Yano tensors of the Taub-NUT space are the following:
\begin{equation}
f_i = 8m(d\chi + \cos\theta d\varphi)\wedge dx_i -
\epsilon_{ijk}(1+\frac{4m}{r}) dx_j \wedge dx_k,~~~~ i,j,k=1,2,3
\end{equation}
\begin{equation}
f_Y = 8m(d\chi + \cos\theta  d\varphi)\wedge dr +
4r(r+2m)(1+\frac{r}{4m})\sin\theta  d\theta \wedge d\varphi
\end{equation}
where the first three are covariantly constant, while the last
one has only one non-vanishing component of the field strength:
\begin{equation}
{f_{Y}}_{r\theta;\varphi} = 2(1+\frac{r}{4m})r\sin\theta.
\end{equation}

Using Eq.(15) we can construct from the Killing-Yano tensors
(16) and (17) the supercharges $Q_i$ and $Q_Y$. The supercharges
$Q_i$ together with $Q_0$ from Eq.(5) realize the N=4
supersymmetry algebra [12]:
\begin{equation}
\left\{ Q_A , Q_B \right\} = -2i\delta_{AB}H~~~,~~~A,B=0,\dots,3
\end{equation}
making manifest the link between the existence of the
Killing-Yano tensors and the hyper- \"ahler geometry of the
Taub-NUT manifold.

Starting from the geometric objects (9), (16) and (17) we shall
express in terms of them all the conserved quantities 
of the geodesic motion in Taub-NUT spinning space.

In order to find the angular momentum (10) and the ``relative
electric charge'' for the spinning case we shall consider the
generalized Killing equation (4) for $n=0$ with the Killing
vectors (9) in the right hand side. For the Killing scalar
${\cal J}_0$ we get [3,12]:
\begin{equation}
B_{A} = \frac i2 R_{A[\mu;\nu]}\psi^\mu \psi^\nu~~~,
~~~A=0,...,3
\end{equation}
and the total angular momentum and the "relative electric
charge" become in the spinning case [14]:
\begin{eqnarray}
{\vec J}&=& {\vec B} + {\vec j}\\
J_0 &=& B_0 + q
\end{eqnarray}
where ${\vec J} =(J_1, J_2, J_3)$ and ${\vec B} =
(B_1, B_2, B_3)$.

The components (21) of the angular momentum satisfy,
as expected, the $SO(3) $algebra:
\begin{equation}
\left\{ J_{i} , J_{j} \right\} = \epsilon_{ijk} J_k~~~,~~~
i,j,k=1,2,3.
\end{equation}

Moreover, the supercharges $Q_i$ transform as vectors at spatial
rotations
\begin{equation}
\{Q_i ,J_j\}=\epsilon_{ijk} Q_k
\end{equation}
while $Q_Y$ and $Q_0$ behave as scalars.

We note also that the bracket of $Q_Y$ with itself can be 
expressed in terms of the hamiltonian, angular momentum 
and ``relative electric charge'':
\begin{equation}
\{Q_Y , Q_Y\}=-2i\left( H +\frac{{\vec J}^{~2} 
- {J_0}^2}{4 m^2}\right).
\end{equation}

On the other hand there are some conserved quantities which are 
peculiar to the spinning case. Taking into account the existence 
of the Killing-Yano covariantly constants tensors $f_i$ (16), 
three constants of motion can be obtained from the homogenous 
part of the generalized Killing equation (4) for $n=0$
\begin{equation}
S_{i} = \frac i4 f_{i\mu\nu}\psi^\mu \psi^\nu
~~~,~~~ i=1,2,3
\end{equation}
which realize an $SO(3)$ Lie-algebra similar to that of the
angular momentum (23):
\begin{equation}
\left\{S_{i} , S_{j}\right\} = \epsilon_{ijk}
S_{k}.
\end{equation}

These components of the spin are separately conserved and can be
combined with the angular momentum $\vec{J}$ to define a new
improved form of the
angular momentum $I_{i} = J_{i} - S_{i}$ with the property that it 
preserves the algebra 
\begin{equation}
\left\{ I_{i},I_{j} \right\} = 
\epsilon_{ijk}I_{k}
\end{equation}
and that it commutes with the $SO(3)$ algebra generated by the
spin $S_{i}$ 
\begin{equation}
\left\lbrace I_{i},S_{j} \right\rbrace = 0.
\end{equation}

To get the spin correction to the Runge-Lenz vector (12) it is
necessary to investigate the generalized Killing equations (4)
for $n=1$ with the Killing tensor ${\vec K}_{\mu\nu}$  (12) in
the right hand side. Unfortunately, it is not possible to find a
closed, analytic expression of the solution of this equation in
terms of the Killing tensor ${\vec K}_{\mu\nu}$ 
and its derivatives. But assuming that the Killing
tensor ${\vec K}_{\mu\nu}$ can be written in terms of
Killing-Yano tensors, the construction of the conserved
Runge-Lenz vector in Taub-NUT spinning space becomes feasible [3].

An explicit evaluation shows that the components of the
Runge-Lenz vector (12) can be written in terms of the
geometrical objects (9), (16), (17):
\begin{equation}
K_{i\mu\nu} = m\left( f_{Y\mu\lambda}
{{f_{i}}^\lambda}_\nu + f_{Y\nu\lambda} {{f_{i}}^\lambda}_\mu
\right) +\frac{1}{8m} (R_{0\mu} R_{i\nu} + R_{0\nu} R_{i\mu}).
\end{equation}
This equation corrects some old formulas from the literature [4].

Starting with this decomposition of the Runge-Lenz vector 
${\vec K}$ from the scalar case, it is possible to express the
corresponding conserved quantity ${\vec {\cal K}}$ in the 
spinning case:
\begin{equation}
{\cal K}_i = 2m \left( -i\{ Q_Y , Q_i\} + \frac{1}{8 m^2} J_i
J_0 \right).
\end{equation}

This expression differs from previous results present in the
literature [3, 12], and the difference has the origin in the
corrected form of relation (30). A detailed expression of the
components ${\cal K}_{i\mu\nu}$ is:
\begin{eqnarray}
{\cal K}_i &=& 2m \left[ \left((f_Y f_i)_{(\mu\nu)} +\frac{1}
{16 m^2} R_{i(\mu} R_{0\nu)}\right) \Pi^\mu \Pi^\nu \nonumber
\right.\\
&+&\left.\biggl( {{f_i}^\lambda}_\beta f_{Y\mu\alpha;\lambda} +
{{f_i}^\lambda}_\mu f_{Y\alpha\beta;\lambda} \right.\biggr.
\nonumber\\
&-& \left.\left.\frac{1}{16 m^2}(R_{i\alpha;\beta} R_{0\mu} 
+ R_{0\alpha\beta} R_{i\mu})\right)S^{\alpha\beta}\Pi^\mu \right.
\nonumber\\
&+&\left.\frac{1}{32 m^2} S^{\alpha\beta} S^{\gamma\delta}
R_{i\alpha;\beta} R_{0\gamma\delta}\right]
\end{eqnarray}
where 
\begin{equation}
S^{\alpha\beta}=-i\psi^\alpha \psi^\beta
\end{equation}
can formally be regarded as the spin-polarization tensor of the
particle [1].

Therefore, in the spinning case, the Runge-Lenz
vector contains additional terms linear and quadratic in the
spin. The presence of a contribution quadratic in the spin,
non-existent in Refs.[12,15], is again related to the term $J_i
J_0$ from Eq.(31).
We would like to emphasize that the contribution of the 
$J_i J_0$ term in Eq.(31) is essential in reproducing the known 
vectorial expression of the Runge-Lenz vector in the scalar case,
and gives the correct Poisson-Dirac bracket between two components 
of the Runge-Lenz vector.

The Dirac brackets involving the Runge-Lenz vector (31) are
(after some algebra):
\begin{eqnarray}
\{ {\cal K}_i , Q_0 \} &=& 0\nonumber\\
\{ {\cal K}_i , J_j\} &=& \epsilon_{ijk} {\cal K}_k\nonumber\\
\{ {\cal K}_i , {\cal K}_j \} =&=& \epsilon_{ijk} J_k \left[
\frac{{J_0}^2}{16 m^2} -2H\right]
\end{eqnarray}
and are similar to those known from the scalar case.
The Runge-Lenz vector $\vec{\cal K}$
together with the total angular momentum $\vec{J}$
generates an $SO(4)$ or $SO(3,1)$ algebra depending upon the sign 
of the quantity $\left(\frac{{J_0}^2}{16 m^2}-2E\right)
\vert~_{\psi^\mu=0}$ is positive or negative.

One ought to mention that the interpretation of $S^{\alpha
\beta}$ defined in Eq.(33) as the spin tensor is corroborated by
studying the electromagnetic interactions of the particle. The
condition for the absence of an intrinsic electric dipole moment
of physical fermions (leptons and quarks) can be satisfied
choosing $Q_0=0$ [1,16]. A more general discussion on the
inclusion of this physical constraint and some explicit solution
of the geodesic motion in Taub-NUT spinning space will be
presented elsewhere .

~

One of the authors (M.V.) would like to thank M.Moshe and
M.S.Marinov for useful discussion on the spinning spaces.


\begin{thebibliography}{99}
\bibitem{1} See e.g. R.H. Rietdijk, Applications of supersymmetric
quantum mechanics, Ph. D. Thesis, Univ. Amsterdam (1992), and 
references therein.
%
\bibitem{2}G.W. Gibbons, R.H. Rietdijk and J.W. van Holten, 
Nucl. Phys. {\bf B404} (1993) 42.                               
%
\bibitem{3}D. Vaman and M. Visinescu, Phys. Rev. {\bf D54} 
(1996) 1398.
%
\bibitem{4}G.W. Gibbons and P.J. Ruback, Phys. Lett. {\bf B188} 
(1987) 226; Commun. Math. Phys. {\bf 115} (1988) 267.
%
\bibitem{5}D.J.Gross and M.J.Perry, {\sl Nucl.Phys.} {\bf B226} 
(1983) 29.
%
\bibitem{6}R.D.Sorkin, {\sl Phys.Rev.Lett.} {\bf 51} (1983) 87.
%
\bibitem{7}N.S.Manton, {\sl Phys.Lett.} {\bf B110} (1982) 54; 
id, {\bf B154} (1985) 397; id, (E) {\bf B157} (1985) 475.
%
\bibitem{8}M.F.Atiyah and  N.Hitchin, {\sl Phys.Lett.} {\bf A107} 
(1985) 21.
%
\bibitem{9}G.W. Gibbons and N.S. Manton, Nucl. Phys. {\bf B274} 
(1986) 183.
%
\bibitem{10}L.Gy. Feher and P.A. Horvathy, Phys. Lett. {\bf B182}
(1987) 183; id, (E) {\bf B188} (1987) 512.
%
\bibitem{11}B. Cordani, L.Gy. Feher and P.A. Horvathy, Phys. Lett.
{\bf B201} (1988) 481.
%
\bibitem{12}J.W.van Holten, {\sl Phys.Lett.} {\bf B342} (1995) 47.
%
\bibitem{13}M. Tanimoto, Nucl. Phys. {\bf B442} (1995) 549.
%
\bibitem{14}M.Visinescu, {\sl Class.Quant.Grav.} {\bf 11} (1994) 
1867.
%
\bibitem{15}M.Visinescu, {\sl Phys.Lett.} {\bf B339} (1994) 28.
%
\bibitem{16}M.S.Marinov, private communication.
%
\end{thebibliography}
\end{document}